\documentclass[10pt, conference]{IEEEtran}

\usepackage{amssymb}
\usepackage{amsmath}
\usepackage{balance}
\usepackage{listings}
\usepackage{booktabs}
\usepackage{ifthen}
\usepackage[utf8]{inputenc}
\usepackage{graphicx}
\usepackage[amsmath]{ntheorem}
\usepackage{tabularx}
\usepackage[hyphens]{url}
\usepackage[table]{xcolor}
\usepackage{xspace}

\lstdefinelanguage{JavaScript}{
	keywords={typeof, new, true, false, catch, function, return, null, catch, switch, var, if, in, while, do, else, case, break},
	keywordstyle=\color{blue}\bfseries,
	ndkeywords={class, export, boolean, throw, implements, import, this},
	ndkeywordstyle=\color{darkgray}\bfseries,
	identifierstyle=\color{black},
	sensitive=false,
	comment=[l]{//},
	morecomment=[s]{/*}{*/},
	commentstyle=\color{purple}\ttfamily,
	stringstyle=\color{red}\ttfamily,
	morestring=[b]',
	morestring=[b]"
}

\definecolor{light-gray}{gray}{0.95}
\definecolor{light-red}{rgb}{1, 0.95, 0.95}
\definecolor{light-green}{rgb}{0.95, 1, 0.95}

\newboolean{showcomments}
\setboolean{showcomments}{false}
\ifthenelse{\boolean{showcomments}}
 { \newcommand{\mynote}[2]{
      \fbox{\bfseries\sffamily\scriptsize#1}
        {\small\textsf{\emph{#2}}}}}
        { \newcommand{\mynote}[2]{}}

\newcommand\tabref[1]{Table~\ref{tbl:#1}\xspace}
\newcommand\figref[1]{Figure~\ref{fig:#1}\xspace}
\newcommand\secref[1]{Section~\ref{sec:#1}\xspace}

\newcommand{\logmd}{\texttt{log.md}\xspace}
\newcommand{\noUnreach}{\texttt{no-unreachable}\xspace}

\begin{document}
	\title{Gamification: a Game Changer for Managing Technical Debt? A Design Study}

	\author{\IEEEauthorblockN{Matthieu Foucault, Margaret-Anne Storey}
	\IEEEauthorblockA{University of Victoria\\
	The CHISEL Group\\
	Victoria, BC, Canada\\
	Email: mfoucault@uvic.ca, mstorey@uvic.ca}
	\and
	\IEEEauthorblockN{Xavier Blanc, Jean-R\'{e}my Falleri}
	\IEEEauthorblockA{Universit\'{e} de Bordeaux\\
	LaBRI -- Software Engineering Group\\
	Bordeaux, France\\
	Email: xblanc@labri.fr, falleri@labri.fr}
	\and
	\IEEEauthorblockN{C\'{e}dric Teyton}
	\IEEEauthorblockA{ProMyze\\
	Bordeaux, France\\
	Email: cedric.teyton@promyze.com}
	}

	\maketitle

\begin{abstract}
\textbf{Context:} Technical debt management is challenging for software engineers due to poor tool support and a lack of knowledge on how to prioritize technical debt repayment and prevention activities. Furthermore, when there is a large backlog of debt, developers often lack the motivation to address it. 
\textbf{Objective:} In this paper, we describe a design study to investigate how gamification can support Technical Debt Management in a large legacy software system of an industrial company. 
Our study leads to a novel tool (named \emph{Themis}) that combines technical debt support, version control, and gamification features.
In addition to gamification features, \emph{Themis} provides suggestions for developers on where to focus their effort, and visualizations for managers to track technical debt activities. 
\textbf{Method:} We describe how \emph{Themis} was refined and validated in an iterative deployment with the company, finally conducting a qualitative study to investigate how the features of \emph{Themis} affect technical debt management behavior.  We consider the impact on both developers and managers. 
\textbf{Results:} Our results show that it achieves increased developer motivation, and supports managers in monitoring and influencing developer behaviors. 
We show how our findings may be transferable to other contexts by proposing guidelines on how to apply gamification.
\textbf{Conclusions:} With this case, gamification appears as a promising solution to help technical debt management, although it needs to be carefully designed and implemented to avoid its possible negative effects.
\end{abstract}

\section{Introduction}

The technical debt (TD) metaphor was first used in 1992 to refer to source code quality issues: \emph{``Shipping first time code is like going into debt. A little debt speeds development so long as it is paid back promptly with a rewrite.''}~\cite{cunningham_wycash_1992} The debt analogy remains relevant more than 25 years later as development cost is one of the main issues facing today's software engineering practitioners~\cite{kruchten_technical_2012}.
Although the TD metaphor applies to several aspects of software developments, including requirements, architecture, and documentation, TD in source code remains the most widely studied~\cite{li_systematic_2015}.
There are various ways to manage technical debt in code, from repayment of TD by rewriting code to make it more maintainable, to preventing TD from occurring in the first place, to monitoring TD to bring awareness of how it changes over time~\cite{li_systematic_2015}.

Technical debt management (TDM) is difficult for software engineers due to poor tool support\footnote{\url{https://insights.sei.cmu.edu/sei_blog/2015/07/a-field-study-of-technical-debt.html}} and a lack of knowledge on how to prioritize TD repayment and prevention activities. But more importantly, it is challenging because many developers lack the mindset and motivation to focus on avoiding or repaying technical debt---many developers consider TD repayment as a time consuming activity that is not guaranteed to provide immediate benefits~\cite{yli-huumo_how_2016}.

TDM can be even more challenging if the project's code base is large and contains a backlog of technical debt from many years of development: our research group was contacted by an industry partner (the French governmental agency P\^{o}le Emploi\footnote{\url{http://www.pole-emploi.org/}}) when they realized their 10-year-old project with more than 550k lines of code contained an extensive backlog of code TD that would require over 2,000 worker-days to fix.
This code TD comes in the form of poor code constructs (also named \emph{code smells}~\cite{beck_bad_1999}) that eventually cause a loss of quality and productivity~\cite{pigoski_practical_1996}.
%
Our industry partner recognized that repaying the TD is essential, but would eventually lead to countless repetitive development tasks without adding any visible value from the developer's perspective.
Similarly, they realized that developers lacked the motivation to avoid adding new TD because adding new \emph{code smells} would have little impact on the already high load of existing debt. 
Thus their problem was twofold:  they needed to motivate developers to both repay existing and avoid adding new TD, and they needed a mechanism that managers could use to track and influence how developers acted towards TDM.

In this paper, we present a problem-driven design study~\cite{sedlmair_design_2012} to address the challenge of managing TD. Our proposed solution to this challenge involved the introduction of \emph{gamification} to address TD. 
By gamification, we refer to the use of game elements in a non-game context~\cite{deterding_game_2011}. We chose gamification as it has already been used in software engineering and shows some promise for improving software processes~\cite{brown_managing_2010,dubois_understanding_2013,biegel_code_2014}, although the empirical evidence of its impact on developer behaviour and motivation is scarce.

Working closely with our industry partner and using a design study methodology (see Section~\ref{sec:Methodology}), we characterized the problem we aimed to address through our research. Then we designed and customized a gamification tool, \emph{Themis}, that integrates with our partner's project version control tool as well as with the SonarQube tool for identifying and measuring TD (see Section~\ref{sec:themis}). \emph{Themis} uses gamified features such as points, leaderboards, and challenges as a way to motivate developers and help managers with TDM.  The tool was designed in an iterative manner (through an early deployment of the tool) in response to the needs elicited from our partner.  

We also studied how \emph{Themis} influences developer and manager behaviours and increases motivation towards managing technical debt by surveying both managers and developers after they used it for three months (see Section~\ref{sec:Validation}). We discovered how and why \emph{Themis} and, in turn, gamification positively influences developer behaviour and motivation as well as how it supports managers. But we also learned that gamification must be treated with care as it may not be suitable for all developers and project contexts. 

To motivate the need for this research, Section~\ref{sec:Background} presents some background on technical debt, describes tools for detecting technical debt, and provides an overview of gamification concepts and how gamification has been applied in software engineering. In Section~\ref{sec:Discussion}, we consider how our findings may be transferable to other developer contexts and suggest guidelines for how gamification may be applied to the task of technical debt management.  
We conclude the paper by making a plea for more research on this topic and present some future research directions (see Section~\ref{sec:conclusion}).  
  

\section{Background}
\label{sec:Background}

We provide background and discuss related work on technical debt and tools 
for measuring 
it. We also introduce the concept of gamification and review how it has been applied in software engineering.

\subsection{Technical Debt}
\label{sub:BackgroundTechdebt}

 
Although the technical debt (TD) metaphor has been used for many years, the formalization of TD is still a work in progress~\cite{siebra_theoretical_2016} and most research efforts studying TD are fairly recent. In Li et al.'s systematic mapping study on technical debt, only four primary studies on the topic of TD were identified between 1992 and 2008, while at least 15 studies have been published each year since 2010~\cite{li_systematic_2015} and a special issue on technical debt was published in the Journal of Systems and Software in 2016~\cite{falessi_introduction_2016}.
 
As with most metaphors, TD is subject to interpretation and its definition can vary.
TD can be related to a wide range of software artifacts including source code, 
requirements, documentation, development process, architecture, and even people (e.g.,\ having software expertise concentrated in too few people)~\cite{alves_towards_2014, li_systematic_2015}.
TD may be created deliberately or inadvertently and it can be reckless or prudent~\cite{fowler_technicaldebtquadrant_2009}.

In this paper, we consider code TD, whether it is deliberately or inadvertently added.
To help managing code TD, developers use tools called \emph{linters} to automatically identify \emph{code smells}~\cite{beck_bad_1999} and pinpoint parts of the code that should be fixed to repay the debt~\cite{curtis_estimating_2012,letouzey_managing_2012}.
The \emph{code smells} considered by these linters are described by rules that can be automatically checked against a given source code thanks to static analysis.
Some examples of linters are PMD\footnote{\url{http://pmd.sourceforge.net/pmd-4.3.0/rules/index.html}} or Checkstyle\footnote{\url{http://checkstyle.sourceforge.net/checks.html}} for Java and ESLint\footnote{\url{http://eslint.org/}} for JavaScript.
For example, in ESLint, the \noUnreach rule detects code that is unreachable (after \texttt{return}, \texttt{throw}, \texttt{break}, and \texttt{continue} statements)
Listing~\ref{lst:eslint} presents a snippet of JavaScript source code highlighting this rule.

\lstset{backgroundcolor=\color{light-red}}
\begin{lstlisting}[caption={Example of invalid code for the \noUnreach rule.},label={lst:eslint}]
function foo() {
	return true;
	console.log("done");
}

function bar() {
	throw new Error("Oops!");
	console.log("done");
}

while(value) {
	break;
	console.log("done");
}
\end{lstlisting}

Li et al.'s study~\cite{li_systematic_2015} identified that technical debt management consists of different kinds of activities: TD identification, TD measurement, TD prioritization, TD monitoring, TD repayment, TD representation/documentation, TD communication, and TD prevention.
Their mapping study showed that the most investigated activities are TD identification (code analysis, dependency analysis) and TD measurement (calculation models, code metrics).
In practice, TD communication is the most commonly used activity as reported by Yli-Huumo et al.~\cite{yli-huumo_how_2016}.
However, they stated that \emph{``the biggest issue with TD communication has been the gap between technical and non-technical stakeholders''}, which emphasizes the need for tools that help teams communicate about the state of a project's technical debt.
Yli-Hummo et al. also found that providing developers with the proper mindset and motivation for TDM is one of the largest challenges as technical debt management and repayment takes time and can be seen as a waste of effort by stakeholders~\cite{yli-huumo_how_2016}.

%

\subsection{Gamification}
\label{sub:BackgroundGamification}


Gamification is mostly defined as \emph{the use of game design elements in non-game contexts}~\cite{deterding_game_2011}.
This definition implies that a gamified application:  1) is a game and has rules defining (at least) player interactions and quantifiable outcomes~\cite{juul_game_2003}; 2) uses game elements such as feedback, reputation, and rank~\cite{reeves_total_2009}; and 3) has a game design that may make use of challenges, time pressure, or levels~\cite{brathwaite_challenges_2008}.



Gamification is a young domain where few theoretical foundations are available~\cite{seaborn_gamification_2015}. 
The emerging theories focus on player motivation, behavior change and engagement, with specific attention paid to the relationship between intrinsic motivation (aligned with the player's inner values) and extrinsic motivation (coming from external factors)~\cite{deci_meta-analytic_1999}.
The objective of gamification is to increase the intrinsic motivation based on extrinsic motivators. For example, the desire to become a better programmer (an intrinsic motivation) may be realized and enhanced through gaining badges (an extrinsic motivator).  
However, care must be taken to ensure that extrinsic motivators do not lead to decreased intrinsic motivation~\cite{deci_meta-analytic_1999}.  

Research on the use of gamification in software engineering is relatively recent and most studies have focused on the design of tools that introduce gamification~\cite{pedreira_gamification_2015}.  
Sheth et al. proposed a framework called HALO to add gamification into a software engineering environment with the objective to enhance productivity~\cite{sheth_halo_2011}. This framework was later used to improve the teaching of software design and testing~\cite{sheth_gameful_2015}. 
Singer and Schneider developed a system using points, badges, and leaderboards to provide an incentive for developers to commit their code more often. After conducting an experiment with 37 students, their interviews showed that the tool increased the participants' awareness of the other developers' activity~\cite{singer_it_2012}.
Steffens et al. developed a preliminary framework of how gamification can be used to support and improve collaboration in software engineering~\cite{steffens_using_2015}, as did Dal Sasso et al., who also proposed a framework to create gamified environments in software engineering~\cite{dal_sasso_how_2017}.
Vasilescu et al. investigated how gamification elements (such as reputation points and badges) enhance social knowledge sharing on and across sites such as GitHub and Stack Overflow~\cite{vasilescu_human_2014}.
Passos et al. explored how different releases of a product can be mapped to gaming levels~\cite{passos_turning_2011}.
Snipes et al. proposed how gamification can be used to improve developers' coding activities (e.g., refactoring)~\cite{snipes_towards_2013}.
Prause et al. conducted a field study of how gamification can promote the creation of Javadoc software documentation in an agile environment~\cite{prause_field_2012}.
LaToza et al. explored how gamification can play a role in crowdsourcing development work~\cite{latoza_crowd_2013}.
Duarte et al. explored how gamification can enhance requirements elicitation~\cite{duarte_collaborative_2012}. 
Gamification is also making its way into mainstream development environments such as Visual Studio\footnote{\url{https://channel9.msdn.com/achievements/visualstudio}}.  

To date, few studies have evaluated gamification while even fewer (as of 2014, only six) have investigated how gamification impacts developer motivation in industrial settings~\cite{pedreira_gamification_2015}.
In particular, we see a lack of theoretical foundations to prescribe and evaluate how gamification can play a role in software development.
However, this is not surprising as its application in this domain is still rather new.

In framing our research, we could not find any studies on how gamification impacts technical debt in practice, although Dubois et al.'s preliminary study showed that gamification for avoiding technical debt (such as \emph{code smells}) seems to motivate students in an educational setting~\cite{dubois_understanding_2013} and a white paper by Cognizant claims that the gamification of Sonar in a project reduced quality costs\footnote{\url{https://www.cognizant.com/InsightsWhitepapers/Using-Gamification-to-Build-a-Passionate-and-Quality-Driven-Software-Development-Team.pdf}}.
These preliminary studies provide some evidence that gamification can help address technical debt, but we lack empirical findings on how it may influence developer behaviour and motivation in an industrial setting.
We next describe our design study where gamification was explored as a solution to manage technical debt.

	\section{Research Methodology}
\label{sec:Methodology}


Applying gamification to reduce or avoid technical debt is a relatively new research direction and there are few insights on how gamification may impact the management of technical debt in software development. 
Given the lack of research in this area, our goal is to answer the following exploratory \emph{research questions}:
\begin{itemize}
	\item RQ1: How does gamification impact \emph{developer} behaviour towards technical debt management?
	\item RQ2: How can \emph{managers} use gamification to help them monitor and drive developers' actions on technical debt?
\end{itemize}

Since our research objective was to design an artifact---a gamification tool---to motivate developers to reduce technical debt, we frame our study using the terminology and structure of a \emph{design study methodology}~\cite{sedlmair_design_2012}.  
In this section, we briefly present what a design study methodology is, introduce our industrial partner and discuss how we worked with them to characterize the problem we aimed to address.

\subsection{Design Study Methodology}

According to Selmair et al.~\cite{sedlmair_design_2012}, a design study must first \emph{characterize the problem} to be solved through a designed artifact (tool)---this step is done in collaboration with the identified users of the tool. The next step is to iteratively \emph{design and implement the artifact} with ongoing input from the users. The \emph{tool design is validated} using empirical methods and then the researchers \emph{reflect} on the design study process as well as consider how the findings may be transferable to other settings. As feedback is gathered iteratively during the creation of the tool or artifact, it is expected that the problem characterization may need to be refined. 



\subsection{Study Partner Company}
\label{sec:Case}

Our study involved a French governmental agency, named P\^{o}le Emploi, that provides financial aid for unemployed people (5.5 million people in February 2017).
P\^{o}le Emploi has 50,000 employees among 900 offices in France and a website that receives over 45 million visits each month.
The business depends on a software platform composed of several applications, maintained daily by 300 developers.
Our study focuses on a central application of the platform, which we will refer to as C-App.
C-App is highly strategic as it has a major financial impact on users.
It was initially deployed in 2006 and now consists of 550k lines of code.
It is a J2EE application (with Java 1.6) that is composed of two main components addressing different but related sets of functionalities. The C-App developers use the Eclipse IDE and other tools that are chosen by the company.
%
The source code is hosted on a large, centralized version control system and each release has its own branch within the repository.
Developers are allowed to commit directly to the repository and they conduct face-to-face code reviews with their peers before committing code.
%

C-App is maintained by a group that includes 1 manager, 2 team leads, and 14 developers (divided into a team of 11 developers and a team of 3 developers).
The group manager is responsible for the whole project---they define the main architecture and govern the group.
Each of the team leads directly supervises one of the two C-App components and manages the corresponding development team.
The project follows a scrum methodology and develops in three-week sprints.
For each sprint, the group manager decides which evolution tasks to address and which incidents to fix.


\subsection{Problem Characterization}

In early December 2015, the C-App manager and team leads used the SonarQube~\cite{_sonarqube_????} quality management platform to identify and measure TD.  
SonarQube revealed that the TD in C-App's code would take an estimated 2,000 worker-days of development to address as there were over 15,000 \emph{code smells} that required fixing. 
The manager then asked the team leads to encourage their developers to reduce TD as an underlying project objective.
No other process steps were defined around TD at that time.
SonarQube was somewhat successful in motivating developers to become aware of and address TD, but the high amount of debt (15K issues) that required fixing was demotivating. The developers also perceived fixing TD to be a boring and unrewarding task.  


At the end of 2015, the company approached our research team because we had helped design a commercial technical debt management tool called \emph{Themis}.  At that time, \emph{Themis} already linked commits with technical debt activities and supported some monitoring at a team level. 
However, we realized, as Pedreira et al. do, that: \emph{``Many software engineering tasks, such as testing and maintenance, are considered somewhat `destructive' and not very appealing; i.e., this type of work is not intrinsically motivating, so specific mechanisms to foster motivation are needed.''} ~\cite{pedreira_gamification_2015}
Therefore, we decided to explore the addition of gamification to \emph{Themis}, anticipating that it may help reduce technical debt by changing developer behaviour and motivation.  
\emph{Themis} (with the addition of gamification) was deployed at the company in April 2016 with a day of training for developers.
A new version was deployed in mid-June to improve the scores computation and to provide some new features requested by the users, who were kept in the loop while developing the gamification of \emph{Themis}.
The following section describes the version of \emph{Themis} currently used by our industry partner.



\section{The Themis Solution}
\label{sec:themis}

To help our C-App industrial partner address their problems with technical debt management, we chose to customize and extend the commercially available \emph{Themis}\footnote{\url{http://www.promyze.com}} technical debt management tool. 
This section discusses the customized version of the tool that we deployed after several iterations of design and feedback with the C-App development group.

As a technical debt management tool, \emph{Themis} combines the information produced by a set of linters with the information contained in the version control system (VCS) to identify whose code is breaking or adhering to the linters' rules. When a violation is detected, \emph{Themis} warns the offending code's author and points them to the problematic code.
The gamification layer we added to \emph{Themis} awards a score to developers based on the code they commit. It uses that score to create friendly competition and incentivize people to better manage their TD.

Below, we explain the core principles behind \emph{Themis} and then we present the gamification layer and describe how managers and developers use it to work with TD.

\subsection{Design Principles}
\label{sub:themis-principles}

\emph{Themis} works together with version control systems and linters, expanding the features they provide by identifying which developers are adhering to or violating rules when they commit code.
It performs an analysis of the commits recorded in the VCS to measure their impact on the rules and to link the commits to their corresponding authors.
A commit can yield several positive or negative \emph{actions}.
A negative action is created for each rule violation that is triggered by a commit 
and a positive action is created for each rule violation that is removed by a commit.
The actions extracted from a commit are assigned to the author of the commit.

As an example, Listing~\ref{lst:lint-error} shows a piece of code committed by Bob that violates the \noUnreach rule (line 4, see Section~\ref{sub:BackgroundTechdebt}).
\emph{Themis} sees that the code contains one negative action targeting the rule and assigns it to Bob.

\lstset{backgroundcolor=\color{light-red}}

\begin{lstlisting}[caption={A snapshot of a JavaScript function. First commit as submitted by Bob.},label={lst:lint-error}, float]
function handleClick(event) {
	// Stop event propagation and default behavior
	return false;
	console.log('Clicked', event.target);
}
\end{lstlisting}

Later, Alice made a commit (see Listing~\ref{lst:lint-fixed}) that removes the aforementioned rule violation.
\emph{Themis} sees that this commit contains one positive action targeting the rule and assigns it to Alice.
\lstset{backgroundcolor=\color{light-green}}
\begin{lstlisting}[caption={A snapshot of a JavaScript function. The second and last commit as submitted by Alice.},label={lst:lint-fixed}, float]
function handleClick(event) {
	console.log('Clicked', event.target);
	// Stop event propagation and default behavior
	return false;
}
\end{lstlisting}

This concept of positive or negative action is at the core of \emph{Themis}' gamification layer.

\subsection{Gamification in \emph{Themis}}

The \emph{Themis} gamification layer provides game elements and rules for ``playing the game'', and has its own game design.
%
\emph{Themis} hosts a contest between developers where the goal is to be ranked as high as possible.
Ranking is determined using a score computed from the actions performed by the developers.
Points are rewarded for each action and the sum of the points defines the score. 
Managers control the number of points scored for a given action, serving as ``game masters''.
In our previous example configuration that assigns $+1$ to positive actions and $-1$ to negative actions, the score would be $1$ for Alice and $-1$ for Bob.
Therefore, Alice would have the best ranking and Bob would be in second position.

\emph{Themis} uses a few key game elements.
Each time a developer performs an action, \emph{Themis} provides \textbf{feedback} so that the developer knows the impact of their actions on the score.
A developer can also use this information to better understand the reasons for a score and possibly improve their coding activities. 
\emph{Themis} also shows developer \textbf{ranking} on a leaderboard.
Scores are visible by everyone and 
a person's ranking is updated each time a new action is performed.
Associating ranking with self-reputation encourages developers to pay better attention to TD and improve their coding.

The game design used by \emph{Themis} consists of a timed contest.
For our study, a contest lasted for the duration of a sprint (3 weeks, see Section~\ref{sec:Case}) and the score was reset each time a new sprint started.
We note that the option of resetting the score was requested by our industrial partner. 
In the first \emph{Themis} iteration we deployed, this feature was not supported. As a consequence, the contest never ended which was a mistake in the game design as it could be demotivating.

\emph{Themis} also provides its users with \textbf{challenges} (c.f., Section~\ref{sub:themis-manager}). 
Managers are able to suggest developers perform or avoid certain actions in a specific time window.
Depending on the outcome of the challenge, a bonus or penalty is awarded to all the developers in the group.

\subsection{Manager Views}
\label{sub:themis-manager}

Managers serve as ``game masters'' and do not participate in \emph{Themis} contests.
They configure the points associated with actions and define the challenges---\emph{action plans}---given to developers.
\emph{Themis} provides a view where managers can observe all actions performed by the developers.
This view also shows which rules are being violated the most and the score of each developer, useful information when configuring the points associated with actions and defining challenges.

\emph{Themis} provides a view for configuring the points associated with actions.
For each action, managers can configure whether the action is positive or negative and include information about the rule that is associated with the action (such as severity, category, or name).
For example, a manager can configure the system so that positive actions award $2$ points and negative actions deduct $2$ points by default.
They can further customize it so that negative actions that target the \logmd rule deduct $5$ points and positive ones award $10$ points.

\begin{figure}
	\begin{center}
		\includegraphics[width=1\columnwidth,keepaspectratio]{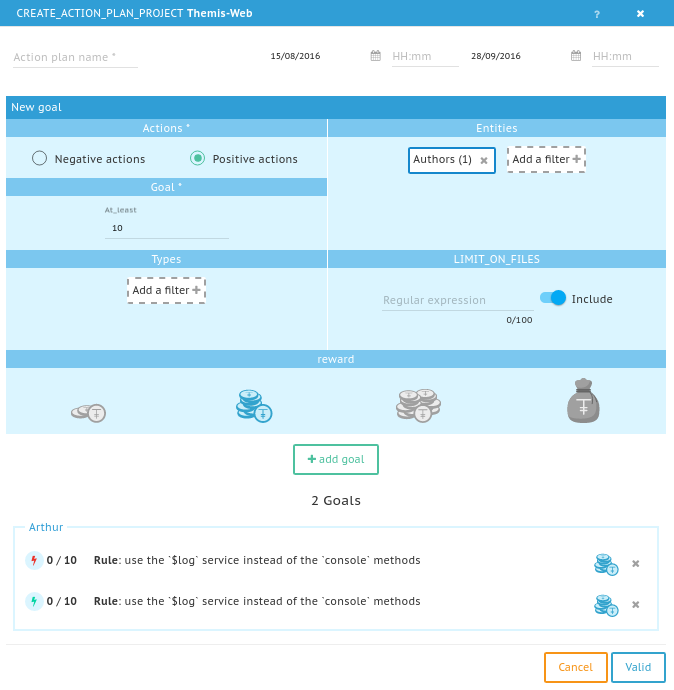}
	\end{center}
	\caption{\emph{Themis} view where managers can set up action plans}
	\label{fig:themis-manager}
\end{figure}

\emph{Themis} provides a view for defining \emph{action plans} to guide developers in managing specific TD.
An action plan is timeboxed, can be assigned to one or more developers, and contains several objectives regarding the actions that should be performed or avoided.
For instance, an action plan could include the following objectives: ``perform less than five negative actions'' and ``perform ten positive actions regarding the \logmd rule''.
An action plan specifies the bonus that will be awarded should the developer succeed or the penalty they will receive should they fail.
When an action plan is assigned to a group, the bonus or penalty is given to all the developers in the group even if only one developer performed the work. The purpose of this is to encourage group collaboration.
Figure~\ref{fig:themis-manager} shows the view where managers can configure action plans.



\subsection{Developer Views}
\label{sub:themis-developer}

\begin{figure}
	\begin{center}
		\includegraphics[width=1\columnwidth,keepaspectratio]{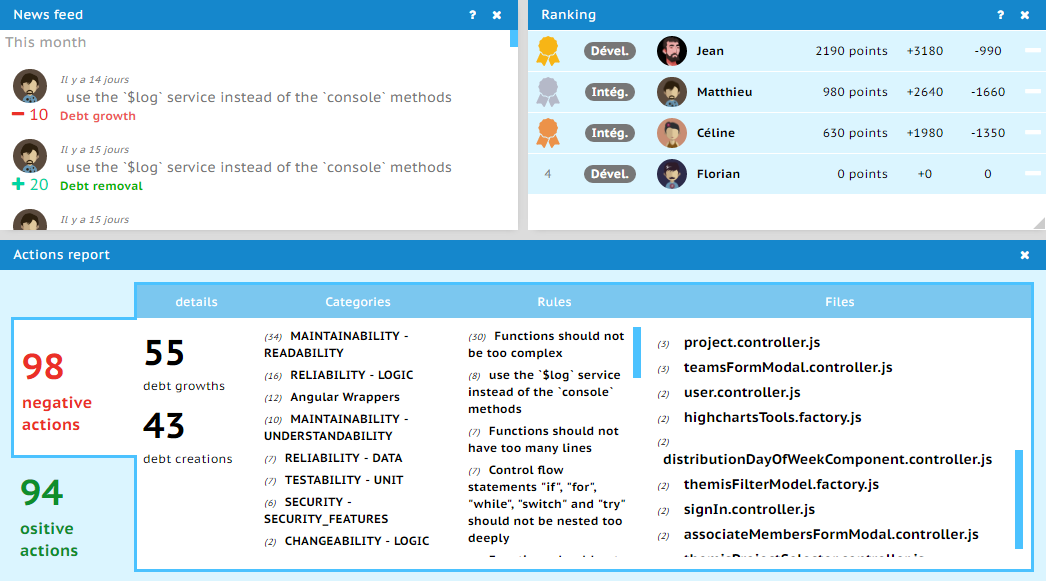}
	\end{center}
	\caption{The \emph{Themis} developer view showing the last actions performed by the developer (in this case, Jean) with their associated score, global ranking, and some more details on the positive and negative actions performed by them.}
	\label{fig:themis-dev}
\end{figure}

Developers participate in contests by performing actions.
\emph{Themis} provides a variety of views (see Figure~\ref{fig:themis-dev}) that developers can use to see their placing within a contest and review the actions they have performed:
\begin{itemize}
	\item A \textbf{newsfeed} that presents the developer's last actions with their associated points and rules. Anonymized versions of other developers' actions are also shown.
	\item A \textbf{leaderboard} that presents the scores and rankings of all developers in the contest.
	\item A \textbf{dashboard} of all the actions performed by the developer with a full description of their associated points, rules, and the files that were affected.
	\item A list of ongoing \textbf{action plans} that have been assigned to the developer by the managers.
\end{itemize}

\begin{figure}
	\begin{center}
		\includegraphics[width=1\columnwidth,keepaspectratio]{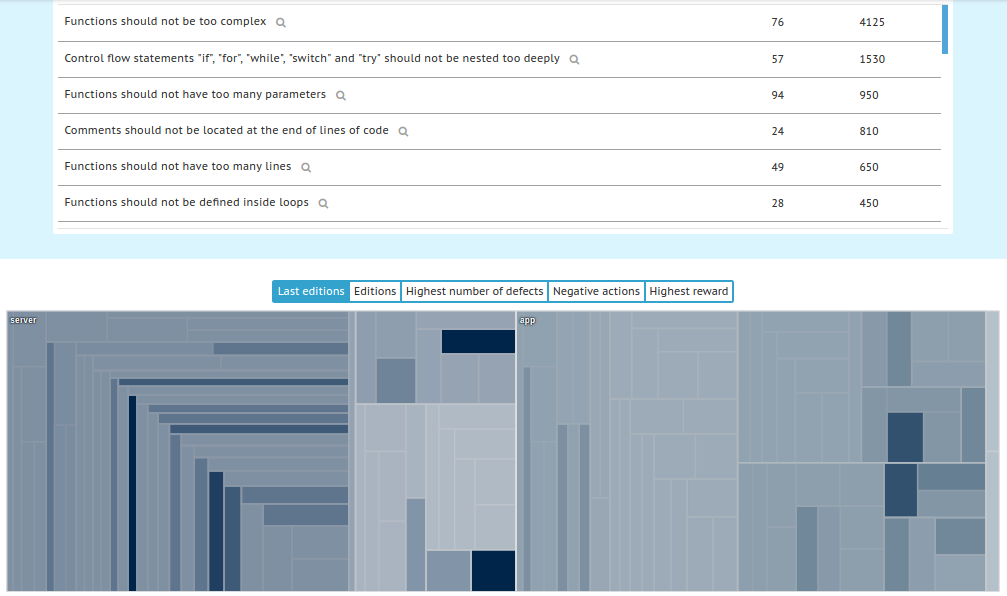}
	\end{center}
	\caption{The \emph{Themis} suggestion module. The top view displays rules in a list ordered by the amount of points that can be rewarded when fixing code. The bottom view is a treemap of the source code.}
	\label{fig:themis-map}
\end{figure}

In addition to these views, \emph{Themis} provides a \textbf{suggestion module} (shown in Figure~\ref{fig:themis-map}) that suggests ways for the developer to score points by pinpointing actions that should be easy for them to perform.
For example, this module identifies and visualizes (using a treemap) any rules that were violated in the files modified by the developer's last commits, or which parts of the source code can provide the highest reward. Developers can use this information to quickly and easily score points, thereby improving their ranking.

	\section{Validation}
\label{sec:Validation}

To validate the final design of the gamified version of \emph{Themis}, we conducted a survey with developers and managers to discern how the tool influenced their behaviours.
We present our data collection and analysis methods, reflect on the limitations of our research process and present our findings.

\subsection{Methods}

We distributed two different survey questionnaires to members of the C-App team 3 months after \emph{Themis} was first deployed: one version of the survey was sent to the manager and both team leads (referred to as managers in the rest of the paper) and another version was sent to the 14 developers.
All 3 managers as well as 8 out of the 14 developers responded to the questionnaires.
All answers were provided anonymously (unless participants chose to give us their email address) and participants were not given any incentive to answer the questionnaire---they generously spent time answering questions without compensation to help us understand the effects of the tool they were using.

The managers' questionnaire included closed- and open-ended questions inquiring about the information they find useful (and why) from the different views available to them, how they use this information, and what feedback they received from their developers.
The developers' questionnaire also focused on the different views provided by \emph{Themis}, asking how (and how often) they use the different features.
Other developer questions were related to how important the score is to them, what steps they take to improve their score, and whether they noticed a change in their motivation to reduce TD since the introduction of \emph{Themis}.
Both questionnaires also asked whether participants think \emph{Themis} had an impact on the TD of their project and what this impact was.

After manually translating the responses from French to English, we used coding to analyze the answers of our questionnaire which consists of labeling data to \emph{``quickly find, pull out, and cluster the segments relating to a particular research
question, hypothesis, construct, or theme.''}~\cite{miles_qualitative_2013}
We performed an initial coding cycle using provisional codes~\cite{miles_qualitative_2013}, i.e., a ``start list'' of codes matching the list of TDM activities described by Li et al. \cite{li_systematic_2015} as well as two codes related to extrinsic and intrinsic motivation, respectively.

To reduce bias during the validation process, we asked two independent researchers to review our survey questions. We also recruited an independent researcher (experienced in qualitative data analysis) to independently review our codes and coding. This independent review led to several iterations of the coding. 
		
Through follow-up questions sent to willing participants, we were able to further verify that the insights we gained (i.e., the main themes) from our analysis of the survey responses resonated with the research participants.  
The answers to these questions helped to confirm our findings from the survey. They also provided additional insights we did not initially probe about in the two sets of questionnaires.

\begin{figure*}
	\centering
	\includegraphics[width=2\columnwidth]{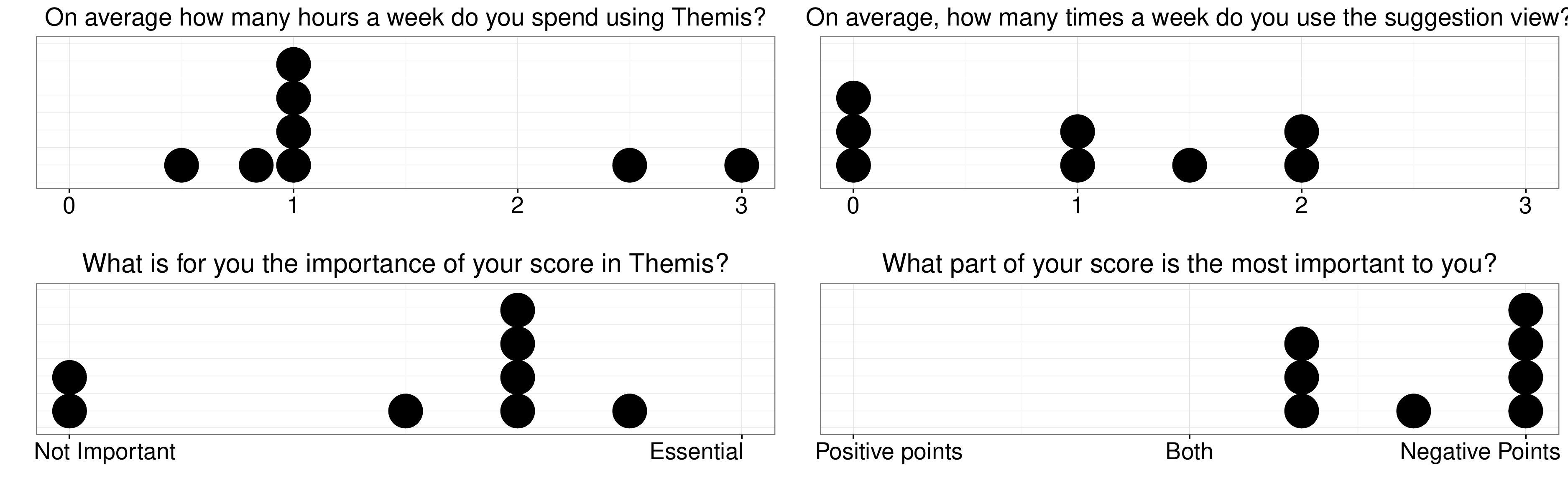}
	\caption{Summary of closed-ended questions from the developers survey. Each circle represents one developer.}
	\label{fig:answers}
\end{figure*}

\subsection{Research Limitations}

There are inevitably a number of limitations with any study, some of which are specific to our chosen research methodology.
Before presenting our findings, we discuss the limitations and the steps we took to offset them.

Throughout our design study, the tool researchers and designers were actively involved in the development and evaluation of the tool.
However, we recognize that this active role of the researcher may have positively influenced the attitudes of the developers and managers towards the tool in the survey and follow-up questions. It even may have changed how they used the tool.
This limitation is an artifact of our research methodology as the role \emph{``of the researcher is central and desirable [to a design study], rather than being a dismaying incursion of subjectivity that is a threat to validity.''}~\cite{sedlmair_design_2012}
Indeed, the close knowledge of the teams and their needs informed the design of the tool so that it would solve their specific problem and it influenced the nature of the questions asked and how they were phrased in the survey and follow-up questions.
This knowledge was also instrumental in the analysis of the responses we received.

To improve the \textbf{credibility} of our findings, we requested \textbf{independent reviewers} to review our survey, codes and themes. Furthermore, we sent follow-up questions to confirm that our findings resonated with the participants. In the presentation of our findings, we report \textbf{discrepant information}, such as when some developers did not find gamification to have any impact on their motivation.

We also note that a design study research methodology aims for \textbf{transferability} rather than reproducibility, as the primary goal is to arrive at a solution that is useful for a specific problem~\cite{sedlmair_design_2012}.
With this in mind, we discuss transferability of our findings in the \secref{Discussion} section of our paper.

We acknowledge that only 8 out of 14 developers answered our survey. Although other developers may have a different experience, our findings are consistent with the managers' point of view of developers' opinion on \emph{Themis}.
To enhance \textbf{traceability}, we provide (in the online supplementary materials for this paper\footnote{\url{https://thechiselgroup.github.io/GamifyTechDebtData/}}) the survey questions (in English and French), the anonymized responses (in English and French) we received, and the final iteration of codes (in English only) that were applied to the responses so that an independent researcher can \textbf{verify}\footnote{although an external reviewer lacking context may find or assign different codes to the data} 
the steps of our analysis or apply a customized version of our instrument to a different case, if desired.
We also provide a copy of the follow-up questions asked of some of the participants 
 as well as the anonymized responses we received.
In anonymizing the responses, we were sensitive to maintaining the confidentiality of the company as well as the confidentiality of individual developers from the two sub-teams (hence we do not identify which of the sub-teams developers belong to in our results).
A copy of our \textbf{ethics approval} is also available.
Finally, we anticipate that the description of the tool given in this paper is sufficiently detailed should other researchers wish to implement and evaluate a similar system.
Additional screenshots from \emph{Themis} are available online\footnote{\url{http://promyze.com/themis}}. 

	\subsection{Findings}

\newcommand\theme[1]{\par\vspace{5pt}\noindent\textbf{#1:}\xspace}
\newcommand\code[1]{\textsc{#1}}

\begin{table*}
	\centering
	\caption{Codes included in the answers of more than half of the respondents. The full list of codes is available in our supplementary online materials. Codes marked with a `*' were present in the list of provisional codes.}
	\label{tbl:codes}
	\begin{tabular}{lp{7cm}r}
		\toprule
		Code & Description & Participants\\ 
		\midrule
		\textsc{TD evolution monitoring} & \emph{Themis} provides a way to monitor actions affecting the evolution of technical debt & 11\\
		\addlinespace[0.25em]
		\textsc{Positive impact on TD reduction} & \emph{Themis} has a positive impact on TD reduction (accelerates TD reduction) & 10\\
		\addlinespace[0.25em]
		\textsc{TD prevention}\textbf{*} & \emph{Themis} helps prevent potential TD from being incurred & 10\\
		\addlinespace[0.25em]
		\textsc{TD repayment}\textbf{*} & \emph{Themis} helps developers resolve existing TD in a software system & 9\\
		\addlinespace[0.25em]
		\textsc{TD prioritization}\textbf{*} & \emph{Themis} ranks identified TD according to certain predefined rules & 7\\
		\addlinespace[0.25em]
		\textsc{Quality standards} & Participants are more attentive to quality standards and processes because of \emph{Themis} & 7\\
		\addlinespace[0.25em]
		\textsc{Limitation: ``one size fits all''} & The tool might not have the same impact on all participants or on other teams & 7\\
		\addlinespace[0.25em]
		\textsc{Extrinsic motivation}\textbf{*} & Participants are motivated by rewards & 7\\
		\addlinespace[0.25em]
		\textsc{Leaderboard} & Participants want to progress up the leaderboard & 7\\
		\addlinespace[0.25em]
		\textsc{TD measurement}\textbf{*} & \emph{Themis} helps quantify and estimate the level of overall TD & 6\\
		\addlinespace[0.25em]
		\textsc{Intrinsic motivation}\textbf{*} & Participants are motivated by their own self-improvement & 6\\
		\bottomrule
	\end{tabular} 
\end{table*}

\begin{table*}
	\centering
	\caption{Themes and participants' adherence}
	\label{tbl:themes}
	\begin{tabular}{p{6cm}cc}
		\toprule
		Theme & Answers fitting the theme & Discrepant answers\\ 
		\midrule
		\emph{Themis} promotes technical debt reduction & $m_1,m_2,m_3,d_1,d_2,d_3,d_4,d_5,d_6,d_7,d_8$ & $m_2,d_7$ \\
		\addlinespace[0.25em]
		\emph{Themis} made developers more attentive to the quality of the code they write & $m_2,m_3,d_1,d_2,d_4,d_5,d_6,d_8$ & $d_3,d_7$\\
		\addlinespace[0.25em]
		Developers appreciate and follow the suggestions provided by \emph{Themis} & $d_1,d_3,d_5,d_6,d_8$ & $d_7$\\				\addlinespace[0.25em]
		Developers follow an opportunistic approach to TD repayment & $d_4,d_6,d_8$ &\\
		\addlinespace[0.25em]
		\emph{Themis} provides monitoring and awareness of TD for individuals and teams & $m_1,m_2,m_3,d_1,d_2,d_3,d_4,d_5,d_6,d_7,d_8$ &\\
		\addlinespace[0.25em]
		\emph{Themis} allows managers to adapt TDM to the context of their projects & $m_1,m_2,m_3$ & $d_3,d_6$ \\
		\addlinespace[0.25em]
		\emph{Themis} promotes ongoing discussion about technical debt between stakeholders & $m_1,m_2,m_3,d_6$ &\\
		\bottomrule
	\end{tabular} 
\end{table*}

Here, we report the findings from the survey and follow-up questions to provide insights on how \emph{Themis}' gamification features impacted developer behaviour and how managers made use of gamification for technical debt management.

Some answers from the closed-ended questions are first summarized in \figref{answers}.
Starting with a list of provisional codes to analyze our survey answers, we established and iterated upon a codebook containing 34 codes.
\tabref{codes} lists the codes that are applicable to more than half of the participants; the full list of codes is available in our online supplementary materials.
Using the codebook, we organized our findings according to seven themes as listed in \tabref{themes}.
We present below our findings, organized according to these themes which are shown in \textbf{bold} while codes are shown in \code{small caps}.
When quoting participants, their id appears as [$m_i$] or [$d_i$] for managers and developers, respectively.

\theme{\emph{Themis} promotes technical debt reduction}
When asked about the overall impact of \emph{Themis} on \code{TD reduction}, all developers and managers agreed that it had a positive impact. Even developers who expressed criticism and did not notice an increase in their own motivation recognized that gamification had a positive impact on other developers. One of the developers commented that \emph{``some seek to be on the top of the leaderboard.''}~[$d_3$]
Another respondent however expressed a contrasting view by stating that \emph{Themis} did not have a significant impact on TD reduction as \emph{``actions to reduce TD were already performed before; there haven't been an increase in actions since.''}~[$d_7$]
Our survey was performed three months after \emph{Themis} was first deployed, which makes TD reduction \emph{``difficult to quantify for the time being.''}~[$m_1$]
Although some data regarding the amount of TD was made available to us and a constant decrease of TD was observed, a statistical analysis of this data would be unreliable due to important confounding factors. The company had migrated their development infrastructure to new servers and operating systems and they implemented configuration changes in Sonar which impacted TD measurement.

\theme{\emph{Themis} made developers more attentive to the quality of the code they write}
Although \emph{Themis} was initially designed to help reduce existing TD backlog and focus on TD repayment, our results show that it had a significant impact on \code{TD prevention} leading to higher \code{quality standards}. One developer mentioned paying \emph{``specific attention every day before committing code''}~[$d_2$] as well using \emph{``quality measurement tools available to [them] before committing [their] code.''}~[$d_6$] 

The main \code{extrinsic} factor that motivates developers is that their points will decrease on the leaderboard if they add TD: \emph{``There is less TD created because it is visible by others through the leaderboard, so we are more careful.''}~[$d_6$]
Moreover, when asked which part of their score they feel is most important, all developers indicated that they focus on having a low number of negative points:  \emph{``You can have 0 points and be a developer applying the `clean code' rules by the book, not generating any defects: that is the goal to reach because ultimately there must not be any TD.''}~[$d_5$]

\code{Intrinsic motivation} stemming from external motivators was expressed by developers as they focus on improving their own code: one developer explicitly told us that he \emph{``mostly look[s] to improve [himself] and having fewer negative points helps with that.''}~[$d_1$]
Another mentioned he uses the leaderboard for \code{self-evaluation}, to \emph{``position [himself] relative to [his] colleagues, not with the goal to show that [he is] better than them, but to see if [he is] as good as them.''}~[$d_6$]

Extrinsic rewards are not effective for all participants in improving intrinsic motivation.
As shown in \figref{answers}, two participants said that their score was not important to them and one of them strongly criticized gamification: 
\emph{``I don’t really like the idea of a leaderboard. [...] I think it is a pity that you need to have a reward (points) to code properly.''}~[$d_3$]
The fact that \code{gamification is not needed for all developers} was confirmed by one manager: \emph{``experienced developers, who are the most sensitive to the non-creation of debt [...] think that \emph{Themis} does not bring anything to them.''}~[$m_3$]

\theme{Developers appreciate and follow the suggestions provided by \emph{Themis}}
Although TD prevention remains the most important in the eyes of participants (see \figref{answers}), \code{TD repayment} was mentioned by all developers and managers as a TDM activity where \emph{Themis} is helpful. Most developers also follow the \code{TD prioritization} of suggestions shown in \emph{Themis}, 
with five out of eight developers using it at least once a week.  
One developer mentioned that he looks at the goals defined by managers \emph{``to see how make quick progress on the leaderboard.''}~[$d_1$]
Developers who are not directly motivated by gamification also \emph{``[use \emph{Themis}] to look for files containing several anomalies to be fixed.''}~[$d_3$]

One developer in contrast indicated that he does not take part in large TD repayment activities and that \emph{``making mass TD repayment actions is absolutely not a part of [his] work. It is uninteresting and there is some development that is way more important that needs to be done.''}~[$d_7$]
He also pointed to a possible side effect of gamification:\emph{``some people spend time fixing classes that are historically not modified only to win a few points when they have pending development/debugging.''}~[$d_7$]
This point was acknowledged by managers who assured us that this view is incidental but they remain careful about it.

\theme{Developers follow an opportunistic approach to TD repayment}
One strategy we observed to reduce TD is what we call \code{opportunistic TD repayment}:
\emph{``[To improve my score in \emph{Themis},] I fix the content of a file that I have to modify for my development. NB: It is not Themis which recommended this file to me (but my development needs), and it is not \emph{Themis} which pointed me to the existing defects in this file (but the quality measurement tools in my IDE).''}~[$d_6$]
For this particular strategy, the initial motivator is clearly the points reward, as this strategy was either mentioned when we asked participants about the steps they take to improve their score, or the participants themselves mentioned the reward as a goal.


\theme{\emph{Themis} provides monitoring and awareness of TD for individuals and the team}
Developers and managers indicated that they use \emph{Themis} to monitor positive and negative actions performed.
Developers mainly use the monitoring features of the main dashboard to \emph{``evaluate [their] work very quickly''}~[$d_6$] and \emph{``see what are [their] areas of improvement''}~[$d_1$]
Managers also strive for \emph{``aggregate information allowing [them] to do medium/long term monitoring.''}~[$m_1$]
They also use \code{monitoring} to update \code{TD prioritization} and to \code{communicate} with their developers:
\emph{``the action reports at the end of the sprint [...] are useful to me in order to create action plans, or to provide reminders to the team or individuals, if needed.''}~[$m_3$]

\theme{\emph{Themis} allows managers to adapt TDM to the context of their project}
Managers emphasized that \emph{``anomalies created and fixed by developers are strongly connected to the project’s context (age, architectural choices, …).''}~[$m_2$], and some rules may not be appropriate to the project or specific modules.
An example shared in the follow up questions was the rule stating that \emph{``cycles between packages should be removed''}, which is tied to initial architectural choices and would now be too costly to remove.
This rule was subsequently disabled in \emph{Themis} by the managers.
Although this \code{intentional TD persistence} is supported by \emph{Themis}, our follow up questions revealed that there are cases where intentional TD could not be ignored by \emph{Themis}.
These cases occur when editing code to comply to a rule---which is not disabled for the whole project---would in turn break another rule, which would require editing one or more different classes.
Furthermore, the project uses non-regression testing further increasing the possible costs of modifying a new class.
In this case, developers would rely on the fact that the score is periodically reset to undo the negative points caused by intentional TD.
Alternatively, one respondent suggested that a manager could undo the negative points someone receives due to intentional debt.

\theme{\emph{Themis} promotes ongoing discussion between stakeholders about technical debt}
While using \emph{Themis}, TD is actively discussed by managers and developers, thus increasing their awareness of TD.
During development phases,
\emph{``as soon as someone sees [newly created TD], they share it loudly in the office to ask the (anonymous) person responsible for it to repay it.''}~[$d_6$]
TD was also discussed extensively during configuration of the tool.
Although configuring the amount of points attributed to each rule can only be done by managers, \code{the whole team was involved in the configuration of \emph{Themis}}.
Thanks to our follow up feedback from one developer, we better understand the process followed by the team to decide on the amount of points to assign to specific rules, and what kind of debates were initiated by \emph{Themis}:
\emph{``Different proposals were made (penalizing more TD creation but leaving repayment points low, penalizing more the creation of more severe defects and reward more the correction of severe defects, etc) and a vote was taken to decide which rule to set up. There are multiple arguments here: some consider that creating TD is worse and has to impact more the developer than the correction of an equivalent TD, others will allow the `right to make a mistake' and consider that a developer who fixed his mistake must be in a neutral state''}~[$d_6$]
The latter was adopted by the team.
This example show that \emph{Themis} puts TD at the center of stakeholders \code{communication}.

	\section{Proposed Guidelines}
\label{sec:Discussion}

\newcounter{gl}
\setcounter{gl}{1}

\newcommand{\guideline}[1]{\vspace{6pt}\noindent{\fbox{\parbox{0.98\columnwidth}
{\emph{Guideline \arabic{gl}:} #1}}}
\addtocounter{gl}{1}
\vspace{-4pt}
}
 
Our research findings lead us to the premise that gamification can play a helpful role in technical debt management, both in reduction and in prevention.  But applying it may not always be wise.  
Here we consider the \emph{transferability} of our work (the primary aim of our study as mentioned earlier) by proposing a set of guidelines that practitioners and researchers may refer to should they wish to apply gamification for managing technical debt.
Although we emphasize gamification for technical debt, we note that many of these principles may apply to other software engineering tasks (such as code review).   Where relevant, we relate the guidelines to the literature, but we remind the reader that there is dearth of theories or advice on how to use gamification for serious work in general~\cite{seaborn_gamification_2015} and even fewer theories for applying gamification in software engineering~\cite{pedreira_gamification_2015}.

\guideline{Nurture a positive team culture} 

The managers and developers from our study were instrumental in setting a positive culture for the use of gamification. In particular, the team did not take ``the game'' too seriously and they enjoyed an atmosphere of \textbf{playfulness}, as one of the developers shared with us how:  \emph{``[The leaderboard] allows to figure out who will bring chocolate croissants at the end of the sprint.''}~[$d_5$] Rather than saying a certain developer was ``last'' they could joke that they owed the team chocolate croissants. 
In particular, we observed that the managers were sensitive to the possible drawbacks of gamification and they were careful not to misuse the information. In turn, the developers trusted their managers and each other.
A trusting culture may not always be possible with a different management style.


\guideline{Tailor gamification to suit different developers}  

Just as different team cultures will influence how gamification impacts behaviour and motivation, the way gamification is used may need to be further adapted to suit different developer characteristics.  As we saw from some of our survey responses, 
\textbf{experienced developers} may not feel the need for the extrinsic motivators, because they do a good job anyway and gamification could force them to use yet another tool on top of the many tools they already use~\cite{tse2016}.  

Developer \textbf{age} and \textbf{gender} may be factors to consider, in our study the developers were all male and between the age of 25-34 and many on the team already played games.  The success of gamification may have been in part due to age, as Dorling et al. note that Generation Y users appreciate clear goals, trackable progress and social rewards~\cite{dorling_gamification_2012}.
Furthermore, younger developers may be more influenced by money, whereas seasoned developers may be more motivated by task variety or challenge~\cite{hall_what_2008}.
In terms of gender, Gneezy et al.~\cite{gneezy_performance_2003} and Vasilescu et al.~\cite{vasilescu_human_2014} found some differences in how females participate on the gamified StackOverflow environment.  We may see differences in how females respond to gamification of technical debt.  
\textbf{Personality} is another consideration as one manager noted: \emph{``Some developers have a more discreet nature and gamification may not be a good motivator for these ones.''}~[$m_2$]
  
\guideline{Adapt the game to project context}

Varied project characteristics may impact which features are needed or should be avoided. 
For example, with \textbf{new projects} where new features are being added at a fast velocity, it may be more important to prevent technical debt: \emph{``it can be less interesting on a new project, since a clean code will not generate any points.''} [$d_8$]
While for \textbf{legacy projects}, repayment may need more motivation:  \emph{``The score is important for older projects with an important debt in order to create a competition and push developers to fix that debt.''}~[$d_8$]

Generation of \textbf{intentional debt}, which we mentioned in our findings, is another example.
There still are cases where developers are punished for TD they have to add to be consistent with existing TD that is too expensive to be repaid. 
When such situations arise, it may be beneficial for managers and developers to \textbf{manually adjust scores} to nurture a feeling of fairness and to maintain morale. 
Currently, this pitfall is mitigated by score resets at the end of every sprint.
This customization was probably essential for the success of \emph{Themis} as previously developers felt overwhelmed as once behind they could not catch up. 

\guideline{Aim for seamless integration with existing processes and tools}

Careful integration of a new tool within a developer's workflow is critical~\cite{johnson_why_2013}. The gamification aspects here are smoothly integrated with the developers' workflow and existing version control and analysis tools.  Furthermore, \emph{Themis} was designed to be easy to use and learn.  These design aspects are critical as Gartner et al. warn about falling prey to this pitfall: \emph{``80\% of the gamified applications will fail to meet their business goals due to a poor design''}.\footnote{\url{http://www.techworld.com/personal-tech/gamification-is-failing-meet-business-objectives-gartner-3425506/}}

\guideline{Keep users in the loop}

Users of a gamified tool should be involved, if possible, in both designing the game and customizing its rules. Keeping everyone involved during the design of the tool is considered an examplar strategy to prevent negative side-effects of gamification, and to make users \emph{commited} to the design of the game~\cite{algashami_strategies_2017}.

\guideline{Monitor how extrinsic motivators influence intrinsic motivation over time}

Tailoring to developer characteristics is important but it is also important to monitor changes in motivation over time, as the game may need further configuration. 
How gamification may impact other management goals over time should also be monitored:  
\emph{``managers must
provide challenging problem-solving tasks,
explicitly recognize quality work, and give
developers autonomy to do their jobs. Managing
these factors effectively will engage
developers and excite them in their work.''}~\cite{hall_what_2008}   

\guideline{Consider when \emph{not} to gamify}

Although gamification shows potential benefits for developers and managers, in agreement with other researchers, we do not advocate that it should be blindly applied as ``lemmingengineering''~\cite{pedreira_gamification_2015}.  There may be other risks to consider---if developers are busy playing the ``game'', what activities does it replace? Will it lead to a lot of code TD repayment, but ignore architectural TD ?
These kinds of strategic issues should be carefully considered.

	\section{Future Work and Conclusions}\label{sec:conclusion}

The use of gamification in software engineering is becoming quite prevalent, in part due to an increased emphasis on data science in software engineering~\cite{kim2016emerging} as well as an increase in the use of social media~\cite{tse2016}.  
But the introduction of some seemingly rather innocent features can have a strong impact\footnote{\url{http://www.hanselman.com/blog/GitHubActivityGuiltAndTheCodersFitBit.aspx}}.  We feel that much more research is needed into the benefits and risks of gamification, while at the same time there are more tasks in software engineering where gamification could be introduced.  

We investigated how gamification could support technical debt management activities, but our study is just the first step in this research. The main outputs from our study are a novel tool design and a set of guidelines and we anticipate that these guidelines can be extended and then used as preliminary propositions in building a \textbf{theory} of the role of gamification for managing technical debt in future research. 

In closing, we make a call for more research studies on gamification. We plan to conduct a \textbf{longitudinal study} of gamification use, which could be very insightful.  Will gamification succeed in reducing TD and supporting TDM activities over one or two years of a project?   
We further agree with Pedreira et al. that there is a need for \textbf{comparative studies}~\cite{pedreira_gamification_2015} and that we should strive to conduct studies of developers doing the same task in a gamified and a non-gamified manner.  However, we note that doing so is very difficult due to many possible confounds---some of which we alluded to in the guidelines.  In the meantime, we hope that our findings from this study will prove useful to both researchers and practitioners interested in the role of gamification in software engineering.



	\vspace{4pt}
	\noindent
	\textbf{Acknowledgements:}
	We thank our research participants and Cassandra Petrachenko for improving our paper.

	\balance{}
	\section{References}
	\bibliographystyle{abbrv}
	\bibliography{bibliography}

\end{document}